\documentclass[12pt,draftclsnofoot,onecolumn]{IEEEtran}

\ifCLASSINFOpdf
\else
\fi
\usepackage{bm}
\usepackage{cite}
\usepackage{amsmath,amsthm, amssymb, latexsym,tikz}
\newcommand{\e}{\mathrm{e}}
\DeclareMathOperator*{\argmin}{arg\,min}  
\DeclareMathOperator*{\argmax}{arg\,max}

\newcommand{\eqdef}{\stackrel{\textrm{\tiny def}}{=}}

\newcommand*{\point}{\makebox[1ex]{\textbf{$\cdot$}}}%

\usepackage[font=small]{caption}

\usepackage{array}
\usepackage[flushleft]{threeparttable}

\begin{document}

\title{Millimeter-Wave Downlink Positioning with a Single-Antenna Receiver}
%
%
% author names and IEEE memberships
% note positions of commas and nonbreaking spaces ( ~ ) LaTeX will not break
% a structure at a ~ so this keeps an author's name from being broken across
% two lines.
% use \thanks{} to gain access to the first footnote area
% a separate \thanks must be used for each paragraph as LaTeX2e's \thanks
% was not built to handle multiple paragraphs
%

\author{\IEEEauthorblockN{Alessio Fascista, Angelo Coluccia, \IEEEmembership{Senior Member, IEEE}, Henk Wymeersch, \IEEEmembership{Member, IEEE}, and Gonzalo Seco-Granados, \IEEEmembership{Senior Member, IEEE}} % <-this % stops a space
\thanks{This work was supported in part by the EU-H2020 project 5GCAR (Fifth Generation Communication Automotive Research and innovation), and in part by the Research and Development Projects of Spanish Ministry of Science, Innovation and Universities TEC2017-89925-R and TEC2017-90808-REDT.}% <-this % stops a space
\thanks{A. Fascista and A. Coluccia are with the Department of Innovation Engineering, Universit\`a del Salento, Via Monteroni, 73100 Lecce, Italy (e-mail: alessio.fascista@unisalento.it; angelo.coluccia@unisalento.it).}
\thanks{G. Seco-Granados is with the Department of Telecommunications and Systems Engineering, Universitat Aut\`onoma de Barcelona, 08193 Barcelona, Spain (e-mail: gonzalo.seco@uab.cat).}
\thanks{H. Wymeersch is with the Department of Electrical Engineering, Chalmers University of Technology, 412 96 Gothenburg, Sweden (e-mail: henkw@chalmers.se).}
}

% note the % following the last \IEEEmembership and also \thanks - 
% these prevent an unwanted space from occurring between the last author name
% and the end of the author line. i.e., if you had this:
% 
% \author{....lastname \thanks{...} \thanks{...} }
%                     ^------------^------------^----Do not want these spaces!
%
% a space would be appended to the last name and could cause every name on that
% line to be shifted left slightly. This is one of those "LaTeX things". For
% instance, "\textbf{A} \textbf{B}" will typeset as "A B" not "AB". To get
% "AB" then you have to do: "\textbf{A}\textbf{B}"
% \thanks is no different in this regard, so shield the last } of each \thanks
% that ends a line with a % and do not let a space in before the next \thanks.
% Spaces after \IEEEmembership other than the last one are OK (and needed) as
% you are supposed to have spaces between the names. For what it is worth,
% this is a minor point as most people would not even notice if the said evil
% space somehow managed to creep in.

% The paper headers
%\markboth{Journal of \LaTeX\ Class Files,~Vol.~14, No.~8, August~2015}%
%{Shell \MakeLowercase{\textit{et al.}}: Bare Demo of IEEEtran.cls for IEEE Journals}
% The only time the second header will appear is for the odd numbered pages
% after the title page when using the twoside option.
% 
% *** Note that you probably will NOT want to include the author's ***
% *** name in the headers of peer review papers.                   ***
% You can use \ifCLASSOPTIONpeerreview for conditional compilation here if
% you desire.

% make the title area
\maketitle

% * <alessio.fascista@unisalento.it> 2018-07-13T10:31:14.635Z:
%
% ^.
\begin{abstract}
The paper addresses the problem of determining the unknown position of a mobile station for a mmWave MISO system.
This setup is motivated by the fact that massive arrays will be initially implemented only on 5G base stations, likely leaving mobile stations with  one antenna.
The maximum likelihood solution to this problem is devised based on the time of flight and angle of departure of received downlink signals. While positioning in the uplink would rely on angle of arrival, it presents scalability limitations that are avoided in the downlink. To circumvent the multidimensional optimization of the optimal joint estimator, we propose two novel approaches amenable to practical implementation thanks to their reduced complexity. A thorough analysis, which includes the derivation of relevant Cram\'er-Rao lower bounds, shows that it is possible to achieve quasi-optimal performance even in presence of few transmissions, low SNRs, and multipath propagation effects.
\end{abstract}

% Note that keywords are not normally used for peerreview papers.

% For peer review papers, you can put extra information on the cover
% page as needed:
% \ifCLASSOPTIONpeerreview
% \begin{center} \bfseries EDICS Category: 3-BBND \end{center}
% \fi
%
% For peerreview papers, this IEEEtran command inserts a page break and
% creates the second title. It will be ignored for other modes.
\IEEEpeerreviewmaketitle

\section{Introduction}

\IEEEPARstart{M}{illimeter-wave} (mmWave) and massive multiple-input multiple-output (MIMO) technologies are currently regarded as strong candidates for next-generation wireless systems, including vehicular and 5G cellular networks. Indeed, such technologies not only are key enabler of high data rates and spectral efficiency \cite{rappaport,lu,larsson,bai,akdeniz}, but also they are promising tools for precise localization thanks to their high temporal resolution and high directivity \cite{taranto,garcia,witrisal,destino,guerra2}.

The theoretical localization performance achievable using mmWave MIMO have been recently investigated in \cite{shahmansoori,shaban,mendrzik}. In \cite{shahmansoori}, the Cram\'{e}r-Rao Lower bound (CRLB) on the position and rotation angle estimates obtained using mmWave from a single transmitter has been derived. Furthermore, a novel position and rotation estimation algorithm based on  compressed sensing that attains the CRLB for average to high signal-to-noise ratio (SNR) is proposed. In \cite{shaban}, fundamental limits of position and orientation estimation for uplink and downlink in 3D-space were presented. Authors in \cite{mendrzik} have shown that non-line-of-sight (NLOS) components can also be exploited to gain additional information for position and orientation estimation.

A few papers have proposed more sophisticated localization schemes, trying to take advantage of the peculiarities of mmWave and (massive) MIMO technologies. In \cite{lin}, a 3D indoor positioning scheme based on hybrid received signal strength (RSS) and Angle-Of-Arrival (AOA), which employs only a single base station (BS) has been presented. A hypothesis testing localization approach is proposed in \cite{deng} exploiting the concept of channel sparsity. A low-complexity AOA-based approach with signal subspace reconstruction is devised in \cite{hu} to localize incoherently distributed sources. An extended Kalman filter (EKF) tracking algorithm that jointly exploits AOA and time of flight (TOF) from uplink reference signals has been proposed in \cite{kovisto}. In \cite{garcia2}, a direct position estimation algorithm is derived. It is based on a compressed sensing framework that exploits some channel properties to identify NLOS signal paths, leading to superior performance compared to other approaches. Authors in \cite{guerra} addressed the problem of positioning based on joint TOF, Angle-of-Departure (AOD), and AOA estimation and investigated the impact of errors in delays and phase shifters. A hybrid time-difference-of-arrival (TDOA), AOA, and AOD localization scheme is presented in \cite{li} based on linearization of a set of local constraints, while in \cite{savic} positioning is addressed using a Gaussian process regressor based on a fingerprinting technique operating on a vector of RSS measurements.

It is worth noticing that almost all the aforementioned approaches heavily rely on the adoption of (possibly large) multi-antenna systems at both transmitter and receiver sides. Although this setup allows for efficient channel estimation by employing high directional beamforming to compensate severe path loss \cite{sayeed}, it requires that commercial massive MIMO implementations for mobile stations (MSs) (e.g., smartphones) will be available in the very near term \cite{cengiz}. However, as recent research showed, it is reasonable to expect that MSs will likely have one (or very few) antennas, while massive arrays will be initially implemented only at the BSs side \cite{gesbert,marzetta}.
%In \cite{zhang}, a Kalman-based tracking algorithm is proposed to track the slow variations of both AOD and AOA. Moreover, a change-detection method is derived to detect abrupt-changes in the channel state. A 2D Unitary ESPRIT-like method is proposed in \cite{miao} to jointly estimate the AODs and AOAs. Good estimation performance could be achieved under high SNR conditions, provided that the channel matrix has been accurately estimated.

In this work, we address the problem of estimating the unknown MS position under a multiple-input single-output (MISO) system setup. The processing is done exclusively on-board at the MS by exploiting the known signals transmitted by a single BS, without any increase in the bandwidth consumption, and requiring an antenna array only on the BS (as opposed to typical MIMO scenarios). While conventional localization schemes mainly focus on AOA estimation, the proposed approach aims at exploiting the AOD of received downlink signals, which can be estimated using a single ominidirectional antenna, thus avoiding the high computational cost required by large arrays. 
%we assume that only a single omnidirectional antenna is available at the MS. 
%Although in the long-term wireless terminals will be equipped with many antennas, it is difficult to believe that a significant deployment of commercial massive MIMO technologies for MSs (e.g., smartphones) will be available in the very near term. 
We show that mmWave and MISO are enabling technologies for designing accurate positioning systems that can be readily implemented in the near future.
%This framework provides a cost-effective solution to design next-generation positioning systems that can be readily implemented in the near future. 
%This apparently simpler setup actually reduces the vector of received data to a single scalar observation, thus posing additional challenges in the location estimation task. Firstly, the AOA cannot be estimated due to the absence of antenna diversity at the receiver side. In addition, the number of observations is not sufficient to allow the estimation of the NLOS channel parameters, hence cannot be harnessed to improve the MS localization accuracy.
%Positioning methods relying on time of flight (TOF), angle of arrival (AOA) and angle of departure (AOD) etc., for MIMO channels has received a considerable attention recently. While several maximum likelihood based and subspace–based algorithms [1], [2], [3] have been proposed to estimate the time delay, AOAs and doppler shift from the received signal at the receive antenna array, there are few results about the AOD estimation in the open literature so far. However the AOD has a critical impact on the transmit correlation matrix of the MIMO channels [4], and it also can be applied to the positioning in nonline–of–sight scenario.

More specifically, the paper provides two kinds of contributions. First, we thoroughly analyze the problem from a theoretical perspective, deriving the exact solution to the Maximum Likelihood (ML) estimation problem and complementing the CRLB-analysis available in the literature \cite{shaban} with a precise assessment of the achievable performance under the considered MISO setup. As a second contribution, we design two novel and practical estimators with reduced complexity. In particular: 
\begin{itemize}
\item a first estimator based on an unstructured transformation of the likelihood function is proposed, which provides an approximate solution to the exact ML problem while avoiding the burden of multidimensional optimization methods;  
\item a second estimation approach based on the method of moments is devised for the case of sufficient amount of received data, which results in a closed-form estimator of the TOF and hence further reduces the complexity to a single one-dimensional search.
\end{itemize}

The numerical analysis in a realistic MISO scenario demonstrates that the proposed estimators can achieve almost the same performance of the exact ML estimator and are able to cope with the different operating conditions at play in typical mmWave channels.

The remaining of the paper is organized as follows. In Sec. \ref{sec::sysmod} we introduce the system model and describe the reference scenario. In Sec. \ref{sec::aodtofpos} we formulate the ML estimation problem and illustrate in details the design and derivation of the proposed low-complexity estimators. Then, in Sec. \ref{sec::FIM}, we derive the fundamental lower bounds on the estimation uncertainty under the considered MISO setup. In Sec. \ref{sec::symanalysis} we analyze the performance, also in comparison with the uplink channel, by means of Monte Carlo simulations in different realistic scenarios. We conclude the paper in Sec. \ref{sec:conclusions}.

\section{System Model}\label{sec::sysmod}
We consider, as a reference scenario, a MISO system with a BS equipped with $N_{\text{BS}}$ antennas and a MS equipped with a single antenna operating at a carrier frequency $f_c$ (corresponding to wavelength $\lambda_c$) and bandwidth $B$. Without loss of generality, the location of the BS is taken in the origin, while we denote by $\bm{p} = [p_x \ p_y]^T$ the unknown position of the MS.
%Locations of the BS and MS are denoted by $\bm{r} = [r_x \ r_y]^T$ and $\bm{p} = [p_x \ p_y]^T$, respectively. In the considered scenario, the value of $\bm{r}$ is assumed to be known, while the MS position $\bm{p}$ is unknown.}Without loss of generality, the location of the BS is taken in the origin, while we denote by .. the unknown position of the MS.

\subsection{Transmitter Model}
We consider the transmission of orthogonal frequency division multiplexing (OFDM) signals, where the BS, implementing a hybrid analog/digital precoder at the transmitting side, communicates with the MS. Particularly, we assume that $G$ signals are transmitted sequentially, where the $g$-th transmission comprises $M$ simultaneously transmitted symbols for each subcarrier $n=0,\ldots,N-1$, i.e.,
\begin{equation}
\bm{x}^{g}[n] = \left[x_1[n]\ \cdots\ x_M[n]\right]^{T} \in \mathbb{C}^{M \times 1} \quad n=0,\ldots,N-1.
\end{equation}
and $P_t = \mathbb{E}\left[\|\bm{x}^{g}[n]\|^2\right]$ the transmitted power with $\mathbb{E}[\point]$ denoting the expectation operator. The symbols are first precoded and then transformed to the time-domain using $N$-point Inverse Fast Fourier Transform (IFFT). A cyclic prefix (CP) of length $T_{\text{CP}} = DT_s$ is added before applying the radio-frequency (RF) precoding, where $D$ is the length of CP in symbols and $T_s = 1/B$ denotes the sampling period. Hereafter, we assume that $T_{\text{CP}}$ exceeds the delay spread of the channel. 

The transmitted signal over subcarrier $n$ at time $g$ can be expressed as $\bm{F}^{g}[n]\bm{x}^{g}[n]$, with $\bm{F}^g[n] \in \mathbb{C}^{N_{\text{BS}} \times M}$ denoting the beamforming matrix applied at the transmitting side. To lower the hardware complexity, in this work we adopt a hybrid beamforming architecture. In particular, assuming that $M^{\tiny \text{RF}}_{\text{BS}}$ RF chains are available at the BS, the beamforming matrix $\bm{F}^g[n]$ can be expressed as 
\begin{equation}
\bm{F}^g[n] = \bm{F}_{\text{RF}}\bm{F}^g_{\text{BB}}[n]
\end{equation}
where $\bm{F}_{\text{RF}} \in \mathbb{C}^{N_{\text{BS}} \times M^{\tiny \text{RF}}_{\text{BS}}}$ is implemented using analog phase shifters with entries of the form $\e^{j\phi_{m,n}}$, where $\{\phi_{m,n}\}$ are given phases, and  $\bm{F}^g_{\text{BB}}[n] \in \mathbb{C}^{M^{\tiny \text{RF}}_{\text{BS}} \times M}$ is the digital beamformer. Furthermore, we impose a total power constraint  $\|\bm{F}_{\text{RF}}\bm{F}^g_{\text{BB}}[n]\|_{\text{F}} = 1$ \cite{Alkhateeb1}. Considering the angular sparsity of the mmWave channels, one usually needs less beams than antenna elements, i.e., $M \leq N_{\text{BS}}$ \cite{brady1,brady2}. It should be noticed that our work does not assume any specific choice of the beamformers $\bm{F}^g[n]$, hence the derived expressions will be completely general. 

\subsection{Channel Model}
We assume that a direct Line-Of-Sight (LOS) link exists between the BS and the MS, and that additional NLOS paths due to local scatterers or reflectors may also be present.
\begin{figure}
\centering
 \includegraphics[width=0.485\textwidth]{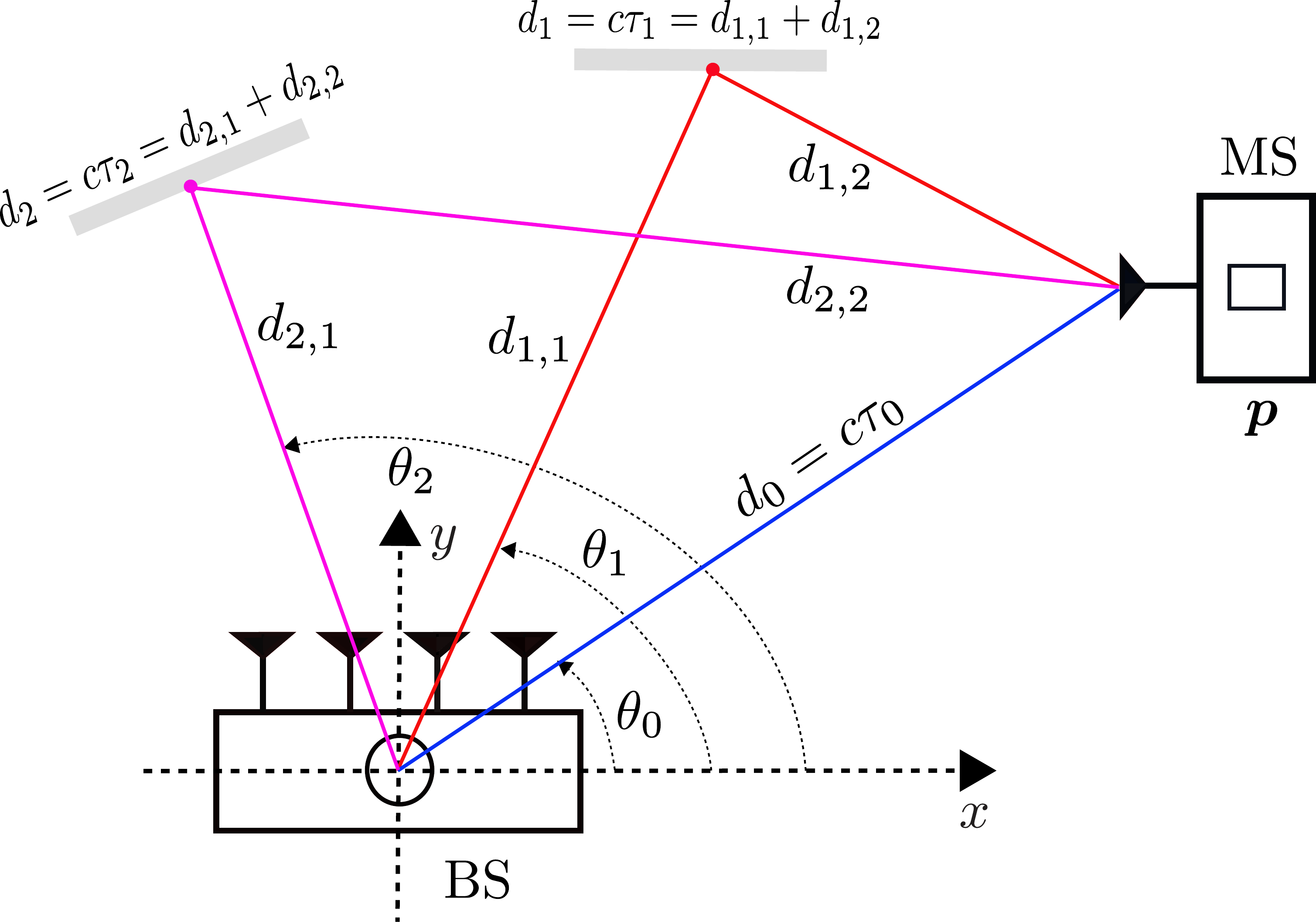}
 	\caption{Geometry of the considered scenario.}
\label{fig:scenario}
 \end{figure}
The different position-related parameters of the channel are depicted in Fig.~\ref{fig:scenario}. These parameters include $\theta_{k}$ and $\tau_k$, denoting the AOD and TOF related to the $k$-th path, respectively. In the following, $k = 0$ corresponds to the LOS link and $k > 0$ denotes the NLOS paths. Considering $P+1$ paths and a constant channel during the transmission of the $G$ signals, the $1 \times N_{\text{BS}}$ complex channel vector associated with subcarrier $n$ can be expressed as 
\begin{equation}\label{eq::fullchannelmodel}
\bm{h}^T[n] = \bm{\Gamma}^T[n]\bm{A}^H_{\text{BS}}
\end{equation}
where we have exploited the fact that $\lambda_n = c/\left(\frac{n}{NT_S} + f_c\right) \approx \lambda_c \,\forall n$ (with $c$ denoting the speed of light), i.e., the typical narrowband condition. Under this model, the array response matrix is given by
\begin{equation}
\bm{A}_{\text{BS}} = [\bm{a}_{\text{BS}}(\theta_0), \ldots,\bm{a}_{\text{BS}}(\theta_P) ]
\end{equation}
% \begin{equation}
% \bm{A}_{\text{Rx}}[n] = [\bm{a}_{\text{Rx},n}(\vartheta_0), \ldots,\bm{a}_{\text{Rx},n}(\vartheta_P) ]
% \end{equation}
and
\begin{equation}
\bm{\Gamma}[n] = \sqrt{N_{\text{BS}}} \times \begin{bmatrix}
    \alpha_0 \e^{\frac{-j2\pi n \tau_0}{N T_s}} \\
     \vdots \\
    \alpha_P \e^{\frac{-j2\pi n \tau_P}{N T_s}} 
  \end{bmatrix}
\end{equation}
% \begin{equation}
% \bm{h}^T[n] = \sqrt{M}\alpha \e^{\frac{-j2\pi n \tau}{NT_S}}\bm{a}^{H}_{\text{Tx},n}(\theta)
% \end{equation}
where $\alpha_p = h_p/\sqrt\rho_p$, with $\rho_p$ the path loss and $h_p$ denoting the complex channel gain of the $p$-th path, respectively.
The structure of the antenna steering vectors $\bm{a}_{\text{BS}}(\theta_p) \in \mathbb{C}^{N_{\text{BS}} \times 1}$ depends on the specific geometry of the considered array. 
Without loss of generality, in the following we consider a Uniform Linear Array (ULA) without mutual antenna coupling and with isotropic antennas, whose steering vector can be expressed as
%\footnote{The response vector  is defined in a similar way, i.e., $\bm{a}_{\text{Rx},n}(\vartheta) = \frac{1}{\sqrt{M_r}} \left[1 \ e^{j\frac{2\pi}{\lambda_n} d\sin\vartheta} \cdots \ e^{j(M_r-1)\frac{2\pi}{\lambda_n}d\sin\vartheta}\right]^T$.} 
\begin{equation}\label{eq::steervect}
\bm{a}_{\text{BS}}(\theta) = \frac{1}{\sqrt{N_{\text{BS}}}} \left[1 \ e^{j\frac{2\pi}{\lambda_c} d\sin\theta} \cdots \ e^{j(N_{\text{BS}}-1)\frac{2\pi}{\lambda_c}d\sin\theta}\right]^T
\end{equation}
%where $\lambda_n = c/\left(\frac{n}{NT_S} + f_c\right)$ is the signal wavelength at the $n$-th subcarrier, $c$ is the speed of light,
where $d = \frac{\lambda_c}{2}$ denotes the ULA interelement spacing. 
%We observe that for $B \ll f_c$, $\lambda_n \approx \lambda_c \forall n$, and \eqref{eq::steervect} reverts to the well-known narrowband model.

\subsection{Received Signal Model}
The received signal related to the $n$-th subcarrier and transmission $g$, after CP removal and Fast Fourier Transform (FFT), is given by
\begin{equation}\label{eq::recsignals}
%\bm{y}^{g}[n] = \bm{H}[n]\bm{F}^{g}[n]\bm{x}^{g}[n] + \bm{\nu}^{g}[n]
y^{g}[n] = \bm{h}^T[n]\bm{F}^{g}[n]\bm{x}^{g}[n] + \nu^{g}[n]
\end{equation}
where $\nu^g[n]$ is the additive circularly complex Gaussian noise with zero mean and variance $\sigma^2$. The ultimate goal of this work is to estimate the unknown MS position $\bm{p}$ from the set of all received signals 
 
%$\mathcal{Y} = \{y^{g}[n], g=1,\ldots, G, n=0,\ldots,N-1\}$. 
\begin{equation}
\bm{Y} = \begin{bmatrix}
    y^1[0] & \cdots & y^G[0] \\
    \vdots& \ddots & \vdots \\
    y^1[N-1] & \cdots & y^G[N-1] 
  \end{bmatrix}.
\end{equation}
To this aim, we  focus on the estimation of the LOS position-related parameters $\theta_0$ and $\tau_0$. Based on such estimates, the unknown MS position $\bm{p}$ can be determined by recalling that the TOF defines a circle centered in the BS and with radius $d_0 = c\tau_0$ from the MS, according to
\begin{equation}\label{eq::TOFpos}
p_x^2 + p_y^2 = d_0^2
\end{equation}
while the AOD is related to the unknown MS position as
\begin{equation}\label{eq::AODpos}
\theta_0 = \mathrm{atan2}(p_y,p_x)
\end{equation}
where the function $\mathrm{atan2}(y,x)$ is the four-quadrant inverse tangent.
Solving \eqref{eq::TOFpos}--\eqref{eq::AODpos} and replacing the actual values with the estimated ones readily provides an estimate of the MS position according to
\begin{equation}
\hat{\bm{p}} = \hat{d}_0\,[\cos \hat{\theta}_0 \ \sin \hat{\theta}_0]^T.
\end{equation}
Fundamental lower bounds on the estimation uncertainty will be derived to evaluate the  performance.

%$\bm{\nu}^{g}[n] \in \mathbb{C}^{M_r \times 1}$ is additive Gaussian noise vector having zero mean and covariance matrix $\bm{R} = \sigma^2 \bm{I}_{M_r}$ with $\sigma^2$ denoting the noise power, and $\bm{I}_{M_r}$ the $M_r \times M_r$ identity matrix. 

\section{Angle of Departure (AOD) and Time Of Flight (TOF) Estimation}\label{sec::aodtofpos}
In this section, we derive novel algorithms for the estimation of the LOS channel parameters $\theta_0$ and $\tau_0$ in presence of the nuisance parameters $\alpha_0$ and $\sigma^2$. 
\subsection{Joint Maximum Likelihood Estimation}\label{subs::JML}
To formulate the estimation problem, we exploit the fact that, under typical mmWave assumptions, all the paths are resolvable in either the time or space domains, and the multipath components are likely uncorrelated with the LOS \cite{shaban}. Since we are interested in estimating the sole LOS position-related parameters, NLOS paths can be omitted from \eqref{eq::fullchannelmodel} and the multipath parameter estimation can be then reduced to a problem of single-path estimation, that is, 
%make use of an unstructured model for the multipath components in \eqref{eq::fullchannelmodel}. This is tantamount to considering each NLOS path as an interference, whose effect is modeled as additive white Gaussian noise.
the channel vector given in \eqref{eq::fullchannelmodel} can be re-defined as
 \begin{equation}
 \bm{h}^T[n] = \sqrt{N_{\text{BS}}}\alpha \, \e^{\frac{-j2\pi n \tau}{NT_S}}\bm{a}^{H}_{\text{BS}}(\theta)
 \end{equation}
where only LOS path is considered from now on. To ease the notation, we introduce $\alpha \eqdef \alpha_0$, $\tau \eqdef \tau_0$, and $\theta \eqdef \theta_0$. Consequently, each received signal $y^{g}[n]$, $1 \leq g \leq G$, $0 \leq n \leq N-1$, can be statistically characterized as 
\begin{equation}
y^{g}[n] \sim \mathcal{CN}(\sqrt{N_{\text{BS}}} \alpha \bar{\bm{h}}^T[n]\bm{F}^{g}[n]\bm{x}^{g}[n], \sigma^2)
\end{equation}
where $\bar{\bm{h}}^T[n] = \e^{\frac{-j2\pi n \tau}{NT_S}}\bm{a}^{H}_{\text{BS}}(\theta)$ and all the parameters are treated as deterministic unknowns, except the transmitted symbols $\bm{x}^{g}[n]$ and the beamforming matrix $\bm{F}^{g}[n]$, which are assumed known to the receiver. More precisely, the whole set of unknowns in $\bm{Y}$ can be arranged as
\begin{equation}
\bm{\varphi} =[\theta\ \tau\ \bm{\psi}]^T
\end{equation}
where $\theta$ and $\tau$ represent the sole parameters of interest, while $\bm{\psi} = [\sigma^2\ \alpha]^T$ denotes the vector of nuisance parameters. The ML estimator of $\theta$ and $\tau$ is  given by
\begin{equation}\label{eq::ML1}
(\hat{\theta},\hat{\tau}) = \argmax_{(\theta, \tau)} \left[\max_{\bm{\psi}} L(\theta, \tau, \bm{\psi})\right]
\end{equation}
where $L(\theta, \tau, \bm{\psi}) \eqdef \log f(\bm{Y}|\theta, \tau, \bm{\psi})$ and $f(\point) $ denotes the probability density function of the observations $\bm{Y}$ given $\bm{\psi}$ and both $\theta$ and $\tau$. From \eqref{eq::ML1} it follows that
\begin{align}\label{eq::loglike_1}
L(\theta,\tau,\bm{\psi}) &= - NG\log(\pi \sigma^2) \nonumber\\
& - \frac{1}{2\sigma^2}\sum_{g=1}^G \|\bm{y}^{g} - \sqrt{N_{\text{BS}}} \alpha \bm{w}^{g}\|^2
\end{align}
where we have denoted by
\begin{equation}
\bm{y}^{g} = \begin{bmatrix} y^{g}[0] \\ \vdots \\ y^{g}[N-1]
\end{bmatrix}
\end{equation}
the $g$-th column of the observation matrix $\bm{Y}$, and
\begin{equation}
\bm{w}^{g} = \begin{bmatrix} \bar{\bm{h}}^T[0]\bm{F}^{g}[0]\bm{x}^{g}[0] \\ \vdots \\ \bar{\bm{h}}^T[N-1]\bm{F}^{g}[N-1]\bm{x}^{g}[N-1]
\end{bmatrix}.
\end{equation}
%and $\bm{f}^{g}[n] = \bm{F}^{g}[n]\bm{x}^{g}[n]$ is a known vector.

We start by observing that the resolution of the ML estimation problem is invariant to the knowledge of $\sigma^2$; in fact, if such a parameter were unknown, it could be estimated as $\hat{\sigma}^2 = \frac{1}{NG}\sum_{g=1}^G \|\bm{y}^{g} - \sqrt{ N_{\text{BS}}} \alpha \bm{w}^{g}\|^2$, leading to the same value of the compressed likelihood as for known $\sigma^2$, i.e. 
\begin{equation}\label{eq::loglike_2}
\tilde{L}(\theta,\tau,\alpha) = \sum_{g=1}^G \|\bm{y}^{g} - \sqrt{N_{\text{BS}}} \alpha \bm{w}^{g}\|^2
\end{equation}
where $\tilde{L}(\theta,\tau,\alpha)$ is the  compressed negative log-likelihood function, and the ML estimator of $\theta$ and $\tau$ reduces to
\begin{equation}\label{eq::ML_2}
(\hat{\theta},\hat{\tau}) = \argmin_{(\theta,\tau)} \left[\min_{\alpha} \tilde{L}(\theta,\tau,\alpha)\right].
\end{equation}
It is a simple matter to observe that the minimization of \eqref{eq::loglike_2} with respect to $\alpha \in \mathbb{C}$ is solved by
\begin{equation}
\hat{\alpha} = \frac{1}{\sqrt{N_{\text{BS}}}} \frac{\sum_{g=1}^G(\bm{w}^{g})^H\bm{y}^{g}}{\sum_{g=1}^G\|\bm{w}^{g}\|^2 }.
\end{equation}
Substituting this maximizing value back in \eqref{eq::loglike_2} leads to 
\begin{equation}\label{eq::loglike_3}
\tilde{L}(\theta,\tau) = \sum_{g=1}^G \left\|\bm{y}^{g} - \left(\frac{\sum_{i=1}^G(\bm{w}^{i}(\theta,\tau))^H\bm{y}^{i}}{\sum_{i=1}^G\|\bm{w}^{i}(\theta,\tau)\|^2 }\right)\bm{w}^{g}(\theta,\tau) \right\|^2
\end{equation}
%with $\bm{P}_{\tilde{\bm{w}}^{g}}^{\perp} = \bm{I}_N - (\tilde{\bm{w}}^{g}\tilde{\bm{w}}^{g H})/\|\tilde{\bm{w}}^{g}\|^2$ denoting the orthogonal projector on the one-dimensional space generated by $\tilde{\bm{w}}^{g}$.
 where we highlight the dependency of $\bm{w}$ on both $\tau$ and $\theta$. As it can be noticed, the function in \eqref{eq::loglike_3} cannot be expressed in terms of any projection matrix; furthermore, it is highly non-linear in both the unknown $\theta$ and $\tau$ and does not admit a closed-form solution.
A possible approach to solve the estimation problem could be based on the adoption of a numerical search algorithm\footnote{Iterative search algorithms like the steepest descent algorithm or the Gauss-Newton method cannot be easily applied since the objective function given in \eqref{eq::loglike_3} exhibits several local minima.}; more precisely, a
two-dimensional grid search can be used for a direct minimization of $\tilde{L}(\theta,\tau)$. To overcome the burden of a multidimensional minimization, in the following we derive two novel suboptimal methods to estimate the TOF $\tau$. In so doing, we can put such an estimated value back in \eqref{eq::loglike_3} and then solve for the unknown $\theta$ in the ML problem by resorting to a simple one-dimensional search. 

It is worth remarking that NLOS paths are not considered (nor exploited) at the design stage. Although their number is limited thanks to the sparsity of the mmWave channel, in the simulation analysis conducted in Sec. \ref{sec::symanalysis}, we will investigate the sensibility of the proposed algorithms to multipath effects according to the LOS-to-Multipath Ratio (LMR), the latter defined as the ratio between the power of the LOS component and the sum of powers of NLOS multipath components.
%However, the algorithms performance will be evaluated in a realistic scenario in which additional NLOS paths are present. 

\subsection{Unstructured ML-based TOF Estimation}
In this subsection, we propose a novel  method for the estimation of the TOF $\tau$. For the sake of exposition, we initially consider the case of single transmission, that is, $G~=~1$. Stacking the observations ${y}[n]$ from \eqref{eq::recsignals}, we obtain
\begin{equation}\label{eq::newrecsignal}
\bm{y} = \sqrt{N_{\text{BS}}}\alpha  \bm{D}(\tau) \bm{\bar{X}} \bm{a}_{\text{BS}}^*(\theta) + \bm{\nu}
\end{equation}
where
\begin{equation}
\bm{D}(\tau) = \begin{bmatrix}
    1 & & \\
    & \ddots & \\
    & & \e^{\frac{-j2\pi (N-1) \tau}{N T_s}} 
  \end{bmatrix}
  \end{equation}
and
\begin{equation}
\bm{\bar{X}}= \begin{bmatrix} (\bm{F}[0]\bm{x}[0])^T \\ \vdots \\ (\bm{F}[N-1]\bm{x}[N-1])^T
\end{bmatrix}.
\end{equation}
To formulate the estimation problem, we make use of an unstructured model for the array steering vector instead of the one parameterized by the AOD, i.e., we introduce the generic vector $\bm{b} = \sqrt{N_{\text{BS}}} \alpha \bm{a}_{\text{BS}}^*(\theta)$. Under this model, \eqref{eq::newrecsignal} can be equivalently rewritten as
\begin{equation}
\bm{y} = \bm{D}(\tau)\bm{\bar{X}}\bm{b} + \bm{\nu}. 
\end{equation}
As it can be noticed, the above expression is no longer an explicit function of $\theta$, but depends only on the TOF $\tau$. Starting from this new model and generalizing to arbitrary values of $G \geq 1$, a ML-based estimator of $\tau$, referred to as Unstructured ML (UML), can be derived as
\begin{equation}\label{eq::unstructured_ML}
\hat{\tau}_{\text{UML}} = \argmin_{\tau} \left[\min_{\bm{b}} \sum_{g=1}^G \|\bm{y}^g - \bm{D}(\tau)\bar{\bm{X}}^g\bm{b} \|^2\right]
\end{equation}
where $\bm{\bar{X}}^g, g=1,\ldots,G$, are known matrices which depend on the transmitted sequences and $\bm{b} \in \mathbb{C}^{N_{\text{BS}} \times 1}$ is treated as an unknown nuisance vector. It is not difficult to show that the inner minimization of \eqref{eq::unstructured_ML} is solved by
\begin{equation}\label{eq::bhat}
\hat{\bm{b}}(\tau) = \bar{\bm{X}}_G^{-1}\sum_{g=1}^G (\bar{\bm{X}}^g)^H\bm{D}^H(\tau)\bm{y}^g
\end{equation}
where $\bar{\bm{X}}_G \eqdef \sum_{g=1}^G (\bar{\bm{X}}^g)^H \bar{\bm{X}}^g$ and the existence of $\bar{\bm{X}}_G^{-1}$ only requires $ N \geq N_{\text{BS}}$. Substituting this maximizing value back in \eqref{eq::unstructured_ML} finally yields
\begin{equation}
\hat{\tau}_{\text{UML}} = \argmin_{\tau} \sum_{g=1}^G \|\bm{y}^g - \bm{D}(\tau)\bar{\bm{X}}^g\hat{\bm{b}}(\tau) \|^2
\end{equation}
which can be solved by resorting to a simple one-dimensional search over the space of $\tau$. Putting the above estimate $\hat{\tau}_{\text{UML}}$ back in \eqref{eq::loglike_3} and solving for the unknown $\theta$ readily provides an approximate solution to the original ML problem, but at the reduced cost of two one-dimensional searches.

\subsection{Moment-based TOF Estimation}
In this section, we further investigate the problem of estimating the unknown $\tau$ when a sufficient number of transmissions $G$ is available.
We start by observing that the elements in \eqref{eq::recsignals} can be equivalently re-arranged as 
\begin{align}
\bm{y}[i] &= \left[y^{1}[i] \cdots y^{G}[i] \right]^T \in \mathbb{C}^{G \times 1} \\
\bm{X}[i] &= \left[\bm{F}^{1}[i]\bm{x}^{1}[i] \cdots \bm{F}^{G}[i]\bm{x}^{G}[i] \right] \in \mathbb{C}^{N_{\text{BS}} \times G} \\
\bm{\nu}[i] &= \left[\nu^{1}[i] \cdots \nu^{G}[i] \right]^T \in \mathbb{C}^{G \times 1} \enspace \forall i=0,\ldots,N-1.
\end{align}
We can now express the collected observations as
\begin{equation}\label{eq::y_i}
\bm{y}[i] = \sqrt{N_{\text{BS}}} \alpha \left(\bar{\bm{h}}^T[i]\bm{X}[i]\right)^T + \bm{\nu}[i] \quad i=0,\ldots,N-1.
\end{equation}
If we assume that $G \geq N_{\text{BS}}$, the known matrices $\bm{X}[i]$ have full row-rank, so that the following transformation can be applied
\begin{equation}\label{eq::pseudoinv}
\underbrace{\bm{y}^T[i]\bm{X}^{+}_i}_{\tilde{\bm{y}}^T_i \in \mathbb{C}^{1 \times N_{\text{BS}}}} = \sqrt{N_{\text{BS}}} \alpha \bar{\bm{h}}^T[i] + \underbrace{\bm{\nu}^T_i\bm{X}^{+}_i}_{\tilde{\bm{\nu}}^T_i \in \mathbb{C}^{1 \times N_{\text{BS}}}}
\end{equation}
where $\bm{X}^{+}_i = \bm{X}^H_i(\bm{X}_i\bm{X}_i^H)^{-1}$ denotes the Moore-Penrose right pseudo-inverse matrix. It can be observed that these new transformed observations are ruled by
\begin{equation}
\tilde{\bm{y}}^T_i \sim \mathcal{CN}_{N_{\text{BS}}}( \sqrt{N_{\text{BS}}} \alpha \bar{\bm{h}}^T[i],\sigma^2 \bm{C}_i) \quad i=0,\ldots,N-1
\end{equation}
where $\bm{C}_i = (\bm{X}^{+}_i)^H\bm{X}^{+}_i$. It is worth noting that, differently from \eqref{eq::y_i}, the elements of each vector $\tilde{\bm{y}}^T_i$ are now correlated. Starting from these new observables, we can further define
\begin{equation}\label{eq::zi}
z_i = \tilde{\bm{y}}^T_{i}\tilde{\bm{y}}^*_{i+1} \quad i=0,\ldots,N-2 
\end{equation}
where each $z_i$ is obtained by multiplying two consecutive $\tilde{\bm{y}}^T_i$, with overlap. This transformation reduces the number of available observations only by one (to $N-1$). 

% (which drops to $N/2$), but avoids correlation among $z_i$s. It should be possible to extend the formulation by considering pairwise differences, which reduces the dataset only by one (to $N-1$) but introduces a tridiagonal correlation to be characterized. 

For the new set of data $\bm{z} = [z_0 \cdots z_{N-1}]^T$, by exploiting the independence of the (transformed) noise vectors, it follows (hereafter we omit the dependency on $\theta$ for simplicity)
\begin{align}
\mathbb{E}[z_i] &= N_{\text{BS}}|\alpha|^2 \bar{\bm{h}}^T[i]\bar{\bm{h}}^*[i+1] \nonumber \\
&= N_{\text{BS}}|\alpha|^2 \e^{\frac{j 2\pi \tau}{N T_s}}\bm{a}^{H}_{\text{BS}}\bm{a}_{\text{BS}} = N_{\text{BS}}|\alpha|^2 \e^{\frac{j 2\pi \tau}{N T_s}}. \label{eq:mean_zi}
\end{align}
%For $\lambda_n \approx \lambda_c \forall n$, \eqref{eq:mean_zi} can be further simplified to
%\begin{equation}
%\E[z_i] = M|\alpha|^2 \e^{\frac{j 2\pi \tau}{N T_s}}.
%\end{equation}
Remarkably, this new expression does not depend on the unknown AOD $\theta$, but it is solely parameterized as function of the TOF $\tau$. Therefore, one can build a method-of-moment (MM) estimator from this expression
\begin{equation}\label{eq::moment_tau}
\hat{\tau}_{\text{MM}} = \frac{N T_s}{2\pi} \arg \left\{\frac{1}{N-2} \sum_{i=0}^{N-2} z_i\right\}
\end{equation}
which results in a closed-form estimator of $\tau$. 
By analyzing the transformed observables $z_i$ in \eqref{eq::zi}, we can derive a useful analogy with the classical frequency estimation theory. Similarly to the Kay's method presented in \cite{Kay}, the approach proposed in \eqref{eq::moment_tau} exploits a one-lag sample autocorrelation function to provide a suboptimal estimator of $\tau$.  Nonetheless, even though this method could provide good estimation performance, an additional improvement can be obtained by considering higher lags in the sample autocorrelation function, as shown in \cite{Fitz}. Based on this result, we derive a more general multi-lag extension of the estimator in \eqref{eq::moment_tau} as
\begin{equation}\label{eq::fritz}
\hat{\tau}_{\text{MM}}(L) = \frac{NT_s}{2\pi q(L)} \sum_{\ell=1}^L \ell \arg\left\{\frac{1}{N-\ell-1} \sum_{i=0}^{N-\ell-1} \tilde{\bm{y}}^{T}_{i}\tilde{\bm{y}}^{*}_{i+\ell}\right\}
\end{equation}
with $L < N$ denoting the number of different lags adopted and $q(L) = \sum_{\ell=1}^L \ell^2 = \frac{L(L+1)(2L+1)}{6}$. As it can be observed, this new estimator weights the argument of the sample autocorrelation function by the lag $\ell$. Moreover, notice that \eqref{eq::fritz} with $L = 1$ is equivalent to the one-lag estimator derived in \eqref{eq::moment_tau}.
Plugging the estimate $\hat{\tau}_{\text{MM}}$ in the ML estimator \eqref{eq::loglike_3} and solving for $\theta$ yields a novel approximate solution to the ML problem, which remarkably only requires a single one-dimensional search.

% {\ed \subsection{MS Position Estimation}
% In this section, we provide a simple and practical algorithm that jointly exploits the estimates of the position-related parameters $\tau$ and $\theta$ to determine the unknown MS position $\bm{p}$. %Fig.~\ref{fig::LOSloc} shows the geometry of the LOS link and highlights the relationship between $\bm{p}$ and both the AOD and TOF information. 
% % \begin{figure}
% % \centering
% %  \includegraphics[width=0.3\textwidth]{localization.pdf}
% %  	\caption{Localization in LOS using TOF (blue segment) and AOD (green arc) information.}
% % \label{fig::LOSloc}
% %  \end{figure}
% As is well-known, the TOF defines a circle centered in the BS and with radius $d = c\tau$ from the MS, according to
% \begin{equation}\label{eq::TOFpos}
% p_x^2 + p_y^2 = d^2
% \end{equation}
% while the AOD is related to the unknown MS position as
% \begin{equation}\label{eq::AODpos}
% \theta = \mathrm{atan2}(p_y,p_x)
% \end{equation}
% where the function $\mathrm{atan2}(y,x)$ is the four-quadrant inverse tangent.
% %and the angle is measured counterclockwise with respect to the $x$-axis, as depicted in Fig.~\ref{fig::LOSloc}.
% Solving \eqref{eq::TOFpos}--\eqref{eq::AODpos} and replacing the actual values with the estimated ones readily provides an estimate of the MS position according to
% \begin{equation}
% \hat{\bm{p}} = \hat{d}\,[\cos \hat{\theta} \ \sin \hat{\theta}]^T.
% \end{equation}
% }

\subsection{Complexity Analysis}
\begin{figure}	
\centering
 \includegraphics[width=0.6\textwidth]{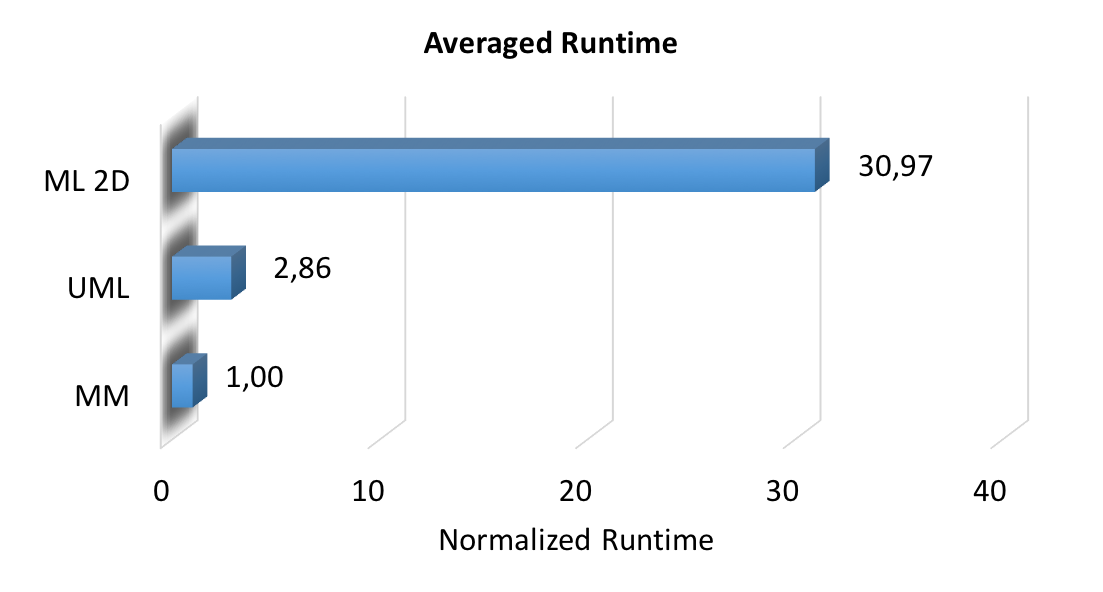}
 	\caption{Averaged runtimes of the proposed estimators.}
\label{fig:runtime}
 \end{figure}

Asymptotically speaking, we observe that the complexity in performing the two-dimensional minimization in \eqref{eq::loglike_3} is on the order of $O(P^2)$, where $P$ denotes the number of evaluation points per dimension, while the two proposed suboptimal approaches only require $O(P)$. However, to perform a precise comparison of the actual complexity for finite values of $P$, we have recorded the runtimes of the estimators executed on the same hardware platform. To conduct the simulation analysis, we consider a grid of $P = 150$ evaluation points and assume $G = 10$. The average runtimes of the estimators normalized by the average runtime of the MM are given in Fig.~ \ref{fig:runtime}. 
As it could be expected, the ML 2D requires by far the longest runtime due to the multidimensional search required for solving \eqref{eq::loglike_3}. On the other hand, the MM has the smallest computational complexity among all the estimators thanks to the closed-form estimation of $\tau$ performed through \eqref{eq::fritz}. The complexity of the UML is only about 3 times larger than that of the MM and, remarkably, is about 10 times lesser than that of the ML 2D. 

\begin{figure} 
\centering
\includegraphics[width=0.5\textwidth]{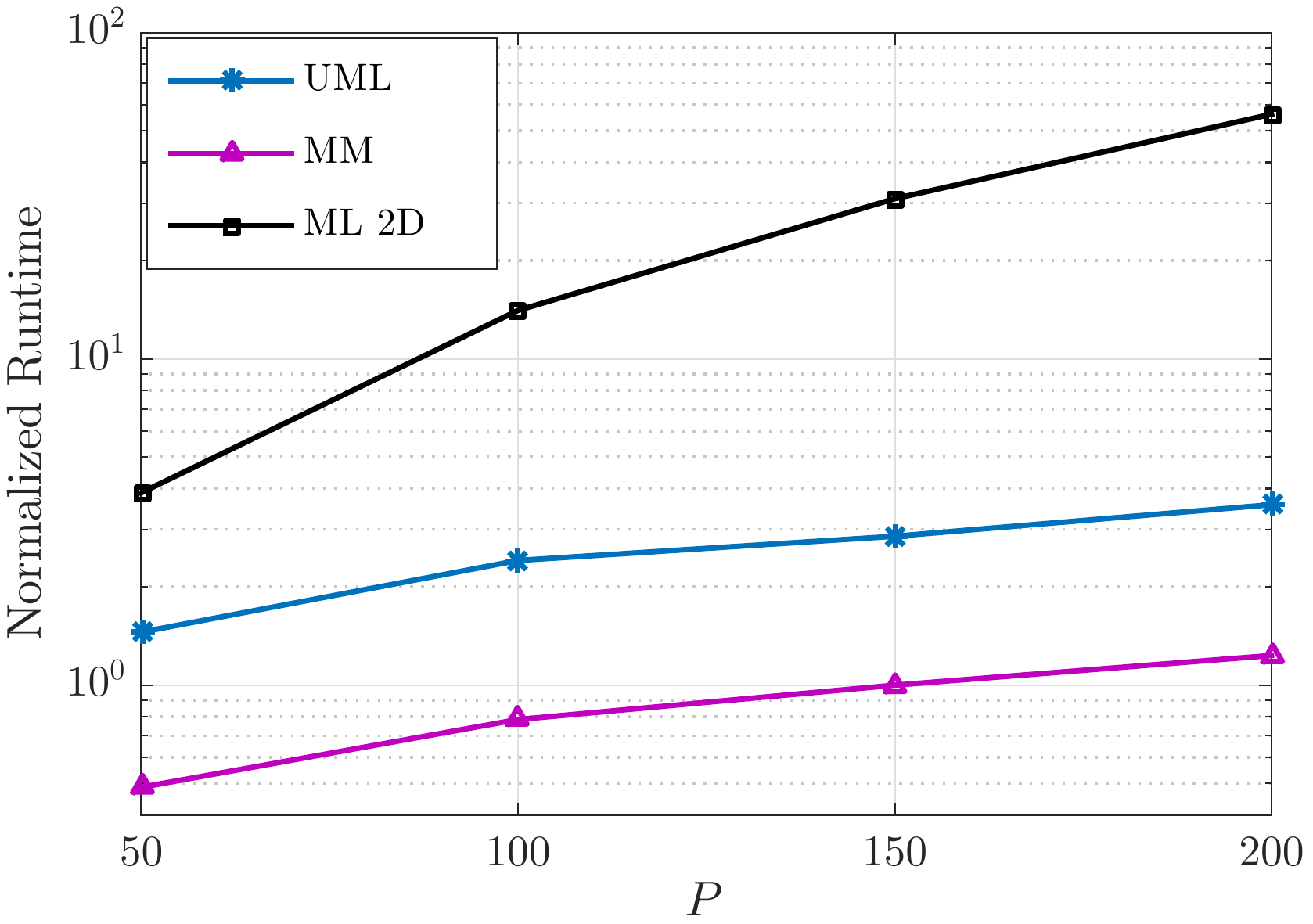}
 	\caption{Runtimes as function of the number of grid points.}
\label{fig:gridcompl}
 \end{figure}

To complete the analysis, we investigate the trend of the computational complexity as a function of the number of grid points $P$. Fig.~\ref{fig:gridcompl} shows the average runtimes for four different values of $P$, normalized by the average runtime of the MM computed for $P = 150$. In agreement with the asymptotic analysis, we observe that both the MM and UML show a roughly linear trend over $P$, although with different slopes. On the other hand, the ML 2D exhibits a superlinear trend, with considerably longer runtimes compared with those of the MM and UML, and a complexity which becomes prohibitive as $P$ increases.

\section{Multiple-Input Single-Output (MISO): Fundamental Bounds}  \label{sec::FIM}
%  Under this setup, the general channel matrix given in \eqref{eq::fullchannelmodel} can be reduced to
%  \begin{equation}
%  \bm{h}^T[n] = \sqrt{M}\alpha \e^{\frac{-j2\pi n \tau}{NT_S}}\bm{a}^{H}_{\text{Tx},n}(\theta)
%  \end{equation}
% where the sole LOS path is considered from now on. To ease the notation, we introduce $\alpha \eqdef \alpha_0$, $\tau \eqdef \tau_0$, $\theta \eqdef \theta_0$, and $M_t \eqdef M$. Consequently, the received signal model in \eqref{eq::recsignals} reduces to
% \begin{equation}\label{eq::recsignals_MISO}
%  {y}^{g}[n] = \bm{h}^T[n]\bm{F}^{g}[n]\bm{x}^{g}[n] + \nu^{g}[n].
% \end{equation}
% where now $\nu^g[n]$ is the additive circularly complex Gaussian noise with zero mean and variance $\sigma^2$. 

In this section, we derive the Fisher Information Matrix (FIM) and the CRLB for the problem of MS position estimation. We start by deriving the bounds on the channel parameters, namely, TOF $\tau$, AOD $\theta$, and path gain $\alpha$. Then, we transform these bounds into the position domain. 
%For the sake of exposition, we consider without loss of generality the case of $G = 1$, i.e., only one transmission is available.
\subsection{FIM Derivation for Channel Parameters}
Let the noise-free observation at subcarrier $n$, transmission $g$
be
\[
m^{g}[n]=\sqrt{M}\alpha\exp\left(\frac{-j2 \pi n\tau}{NT_s}\right)\bm{a}^{H}(\theta)\bm{s}^{g}[n]
\]
where $\bm{s}^{g}[n]=\bm{F}^{g}[n]\bm{x}^{g}[n]$, $\alpha= h/\sqrt\rho \eqdef r\exp(j\phi)$ with $r$ and $\phi$ modulus and phase of the complex amplitude $\alpha$, respectively. Let 
$\bm{\gamma} \in \mathbb{R}^{4 \times 1}$ denotes the vector of the unknown channel parameters
\begin{equation}
\bm{\gamma} = [r\ \phi\ \tau\ \theta]^T
\end{equation}
where we assume without loss of generality that the noise variance $\sigma^2$ is known\footnote{We recall that if $\sigma^2$ were unknown, it could be estimated using the equation provided above \eqref{eq::loglike_2}.}.
Defining $\hat{\bm{\gamma}}$ as an unbiased estimator of $\bm{\gamma}$, it is well-known that the mean squared error (MSE) is lower bounded as \cite{kay}
\begin{equation}
\mathbb{E}_{\bm{Y}|\bm{\gamma}}\left[(\hat{\bm{\gamma}} - \bm{\gamma})(\hat{\bm{\gamma}} - \bm{\gamma})^T \right] \succeq \bm{J}^{-1}_{\bm{\gamma}}
\end{equation}
where $\mathbb{E}_{\bm{Y}|\bm{\gamma}}[\point]$ denotes the expectation parameterized as function of the unknown vector $\bm{\gamma}$ and $\bm{J}_{\bm{\gamma}}$ is the $4 \times 4$ FIM defined as \cite{poor}
\begin{align}
\bm{J}_{\bm{\gamma}} &= \mathbb{E}_{\bm{Y}|\bm{\gamma}}\left[-\frac{\partial^2 \log f(\bm{Y}|\bm{\gamma})}{\partial \bm{\gamma}\partial \bm{\gamma}^T} \right] \nonumber \\
&= \frac{2}{\sigma^{2}}\sum_{g=1}^{G}\sum_{n=0}^{N-1}\Re\left\{ \nabla m^{g}[n]\nabla^{\text{H}}m^{g}[n]\right\} \label{eq::FIM}
\end{align}
where $f(\bm{Y}|\bm{\gamma})$ is the conditional likelihood function of $\bm{Y}$ given $\bm{\gamma}$, while $\nabla m^{g}[n]$ is the gradient of $m$ with respect to $\bm{\gamma}$ given by
\[\!\!\!\!\!
\nabla m^{g}[n]=\left[\!\!\! \begin{array}{c}
\sqrt{N_{\text{BS}}}\exp(j\phi)\exp\left(\frac{-j2 \pi n\tau}{NT_s}\right)\bm{a}^{H}(\theta)\bm{s}^{g}[n]\\
j\sqrt{N_{\text{BS}}}\alpha\exp\left(\frac{-j2 \pi n\tau}{NT_s}\right)\bm{a}^{H}(\theta)\bm{s}^{g}[n]\\
\frac{-j2 \pi n}{NT_s} \sqrt{N_{\text{BS}}} \alpha\exp\left(\frac{-j2 \pi n\tau}{NT_s}\right)\bm{a}^{H}(\theta)\bm{s}^{g}[n]\\
\frac{-j2 \pi}{\lambda_c} d  \cos\theta \sqrt{ N_{\text{BS}}}\alpha\exp\left(\frac{-j2 \pi n\tau}{NT_s}\right)\bm{a}^{H}(\theta)\bm{B}\bm{s}^{g}[n]
\end{array}\!\!\! \right]
\]
with $\Re\{\point\}$ denoting the real-part operator and $\bm{B}=\mathrm{diag}[0\ 1\ \cdots\ (N_{\text{BS}}-1)]$, where $\mathrm{diag}(\point)$ construct a diagonal matrix with its entries. 
We can show that for the matrix $\bm{J}_{\bm{\gamma}}$ to be non-singular when $G=1$,  we need at least two subcarriers and send different pilot sequences on each subcarrier. Similarly, when $N=1$, we need at least two transmissions with different pilot sequences. Further details are found in the Appendix. 
% $N>1$ and $\bm{s}^{g}[n]\neq \bm{s}^{g}[n']$
% the following conditions must be satisfied:
% \begin{enumerate}
% \item $N>1$ 
% \item $G>0$
% \item $\exists n,g:\bm{s}^{g}[n]\neq u\bm{1}$, for some constant
% $u$ and $\bm{1}$ being the all one vector. 
% \end{enumerate}

\subsection{FIM Derivation for Position}
In this section, we derive the FIM in the position domain by applying a transformation of variables from the vector of channel parameters $\bm{\gamma}$ to the vector of location parameters
\begin{equation}
\bm{\eta} = [r\ \phi\ p_x \ p_y]^T.
\end{equation}
The FIM of $\bm{\eta}$ is obtained by means of the $4 \times 4$ transformation matrix $\bm{T}$ as
\begin{equation}\label{eq::FIMpos}
\bm{J}_{\bm{\eta}} = \bm{T}\bm{J}_{\bm{\gamma}}\bm{T}^T
\end{equation}
where 
\begin{equation}
\bm{T} \eqdef \frac{\partial \bm{\gamma}^T}{\partial \bm{\eta}} = \begin{bmatrix}
    \partial r/\partial r & \partial \phi/\partial r & \partial \tau/\partial r & \partial \theta/\partial r \\
    \partial r/\partial \phi & \partial \phi/\partial \phi & \partial \tau/\partial \phi & \partial \theta/\partial \phi  \\
    \partial r/\partial p_x & \partial \phi/\partial p_x & \partial \tau/\partial p_x & \partial \theta/\partial p_x \\
    \partial r/\partial p_y & \partial \phi/\partial p_y & \partial \tau/\partial p_y & \partial \theta/\partial p_y
  \end{bmatrix}.
\end{equation}
The entries of the matrix $\bm{T}$ can be obtained from the relations between the parameters in $\bm{\gamma}$ and $\bm{\eta}$, as expressed in \eqref{eq::TOFpos}--\eqref{eq::AODpos}. More precisely, we have
$$
\partial r/\partial r = \partial \phi/\partial \phi = 1,
$$
$$
\partial \tau/\partial p_x = \frac{p_x}{(p^2_x + p^2_y)^{-\frac{1}{2}}}, \qquad \partial \tau/\partial p_y = \frac{-p_y/p^2_x}{1 + (p_y/p_x)^2},
$$
$$
\partial \theta/\partial p_x = \frac{p_y}{(p^2_x + p^2_y)^{-\frac{1}{2}}}, \qquad \partial \theta/\partial p_y = \frac{1/p_x}{1 + (p_y/p_x)^2},
$$
and the rest of the entries in $\bm{T}$ are zero.

\subsection{Bounds on Position Estimation Error}
The position error bound (PEB) can be readily derived by inverting the FIM $\bm{J}_{\bm{\eta}}$ given in \eqref{eq::FIMpos}, then adding the diagonal entries of the lower right $2 \times 2$ sub-matrix, and taking the square root as
\begin{equation}\label{eq::PEB}
\text{PEB} = \sqrt{\left[(\bm{J}_{\bm{\eta}})^{-1}\right]_{3,3} + \left[(\bm{J}_{\bm{\eta}})^{-1}\right]_{4,4}}
\end{equation}
where the operator $[\point]_{j,j}$ denotes the selection of the $j$-th diagonal entry of $\bm{J}^{-1}_{\bm{\eta}}$.
% \subsection{Uplink Model}
% Let the noise-free observation at subcarrier $n$, transmission $g$ in the uplink channel be
% \begin{equation}
% \bm{y}^g[n] = \sqrt{M}\alpha\exp(-j2 \pi n\tau/NT_s)\mathbf{a}(\theta)x^{g}[n]
% \end{equation}
% with $x^g[n]$ denoting the symbol transmitted at subcarrier $n$. The gradient
% with respect to $[r\ \phi\ \tau\ \theta]^T$ is given by 
% \[\!\!\!\!\!
% \nabla \bm{y}^{g}[n]=\left[\begin{array}{c}
% \sqrt M\exp(j\phi)\exp\left(\frac{-j2 \pi n\tau}{NT_s}\right)\mathbf{a}(\theta)x^{g}[n]\\
% j\sqrt M\alpha\exp\left(\frac{-j2 \pi n\tau}{NT_s}\right)\mathbf{a}(\theta)x^{g}[n]\\
% \frac{-j2 \pi n}{NT_s} \sqrt M \alpha\exp\left(\frac{-j2 \pi n\tau}{NT_s}\right)\mathbf{a}(\theta)x^{g}[n]\\
% \frac{-j2 \pi}{\lambda_n} d  \cos\theta \sqrt M\alpha\exp\left(\frac{-j2 \pi n\tau}{NT_s}\right)\mathbf{D}\mathbf{a}(\theta)x^{g}[n]
% \end{array}\right]
% \]

\section{Simulation Model and Results}\label{sec::symanalysis}
In this section, we present simulation results to evaluate the performance of the proposed estimators in comparison with the two-dimensional ML, as well as the theoretical bounds derived based on the FIM analysis conducted in Sec. \ref{sec::FIM}.

\subsection{Reference scenario}
We consider a scenario representative of outdoor localization in a small open area with maximum distance between BS and MS of 100 meters. The BS is located at the origin of the considered Cartesian reference system and is equipped with $N_{\text{BS}} = 10$ antennas.
%In Table~\ref{tab::par}, we report the actual values of $\tau$ and $\theta$ as function of the distance from the BS. 
As concerns the transmitted signal, we assume a carrier frequency $f_c = 60$ GHz, a bandwidth $B = 40$ MHz, a transmitted power $P_t = 1$ W, and a number of subcarriers $N = 20$ \cite{Maltsev}. The number of simultaneously transmitted beams is $M = 1$ and we vary the number of sequentially transmitted signals $G$ between $1$ and $20$, respectively.

The channel path loss $\rho$ is computed according to \cite{Li1,Li2}. For the specific case of the LOS link, we obtain
\begin{equation}
1/\rho = \mu^2(d_0) \left(\frac{\lambda_c}{4 \pi d_0} \right)^2
\end{equation}
where $\mu^2(d_0)$ denotes the atmospheric attenuation at a distance $d_0$ and the last term is the free space path loss at a distance $d_0$. Following \cite{rappaport}, the atmospheric attenuation $\mu^2(d_0)$ is set to 16 dB/Km. As to the complex channel gain, it can be expressed in terms of $h = a \e^{j \varphi}$ with $a = \sqrt{P_t}$ the amplitude and $\varphi$ the related phase, respectively.

To complete the analysis, we show the performance achievable in the uplink channel and compare them with the ones obtained in the downlink. Notice that, in the uplink case, the model of the received signal is different from the one provided in \eqref{eq::recsignals} since an antenna array should be considered at the receiver side. As a result, the localization will be based on the estimation of the AOA, as typical in many existing approaches, while we recall that the present contribution focuses on leveraging AOD information. Therefore, in the following we will consider the proper modified expressions for the theoretical bounds, as derived in the literature \cite{shaban}. 

 As anticipated in Sec. \ref{subs::JML}, we also investigate the performance of the proposed estimators in presence of multipath propagation. Notice that, in this case, the estimation performance are evaluated in a simulation environment that is not matched to the design assumption of the proposed algorithms. More precisely, it is assumed that, in addition to the direct LOS link, two different NLOS paths are present at the receiver side. Assuming that only one dominant reflector is present in each NLOS path \cite{Li2}, we can compute the path loss $\rho_k$ for the $k$-th NLOS link with path length $d_k$  according to
\begin{equation}
1/\rho_k = \sigma_0^2\mathbb{P}_0(d_k) \left(\frac{\lambda_c}{4 \pi d_k} \right)^2
\end{equation}
where $\mathbb{P}_0(d_k) = (\gamma_r d_k)^2\e^{-\gamma_r d_k}$ denotes the Poisson distribution of the environment geometry with density $\gamma_r$, and $k=1,2$. According to \cite{Li2}, we consider a density $\gamma_r = 1/7$ and set the average reflection loss for the first-order reflection $\sigma_0^2$ to −10 dB with the root-mean-square (RMS) deviation
equal to 4 dB. To analyze the sensibility of the proposed estimators to multipath effects, we consider the following definition of LMR: 
\begin{equation}
\text{LMR} = \frac{P_{\text{\tiny LOS}}}{\sum_{k=1}^2 P^k_{\text{\tiny NLOS}}} = \frac{1/\rho}{1/\rho_1 + 1/\rho_2}
\end{equation}
where $P_{\text{\tiny LOS}}$ is the power associated to the LOS path, while $P^k_{\text{\tiny NLOS}}$ is the power of the $k$-th multipath component.

The elements of the analog beamformers $\bm{F}_{\text{RF}}$ are generated as uniform random values on the unit circle. As concerns the sequences $\bar{\mathbf{x}}^{g}[n]=\mathbf{F}_{\text{BB}}^{g}[n]\mathbf{x}^{g}[n]$, they are computed as complex exponential terms $\e^{j\phi_{n,g}}$ having random phases uniformly distributed in $[0,2\pi)$ along different subcarriers $n$ and different transmissions $g$, respectively.
Finally, we define the SNR in dB as 
\begin{equation}
\text{SNR} \eqdef 10\log_{10}\left(\frac{P_t}{\rho N_0B}\right)
\end{equation}
where $\log_{10}(\point)$ denotes the base-10 logarithm and $N_0 B$ is the receiver noise figure, i.e., $N_0B = k_BT_0B$, $k_B$ being  the Boltzmann constant and $T_0$ the standard thermal noise temperature.

We consider the Root Mean Squared Error (RMSE) as metric to assess the algorithms performance, estimated based on 1000 Monte Carlo trials. 
\subsection{AOD, TOF and Position Estimation in LOS}\label{subs::LOS}

%  \begin{table}
%  \centering 
%  \caption{Values of the actual $\tau$, $\theta$, and SNR for different distances from the BS.}\label{tab::par}
%  \begin{tabular}{ccccccc}
% \hline%
%  \multicolumn{1}{c}{distance (m)} & $60$& $50$ & $40$ & 30 & 20 & 10\\
%  \hline%
%  SNR (dB) & 18 & 19 & 20 & 22 & 24 & 27 \\ \hline 
%  \end{tabular}
%  \end{table}

In this section, we assess the performance of the proposed algorithms assuming only the LOS component is present. 
\begin{figure}	
\centering
 \includegraphics[width=0.5\textwidth]{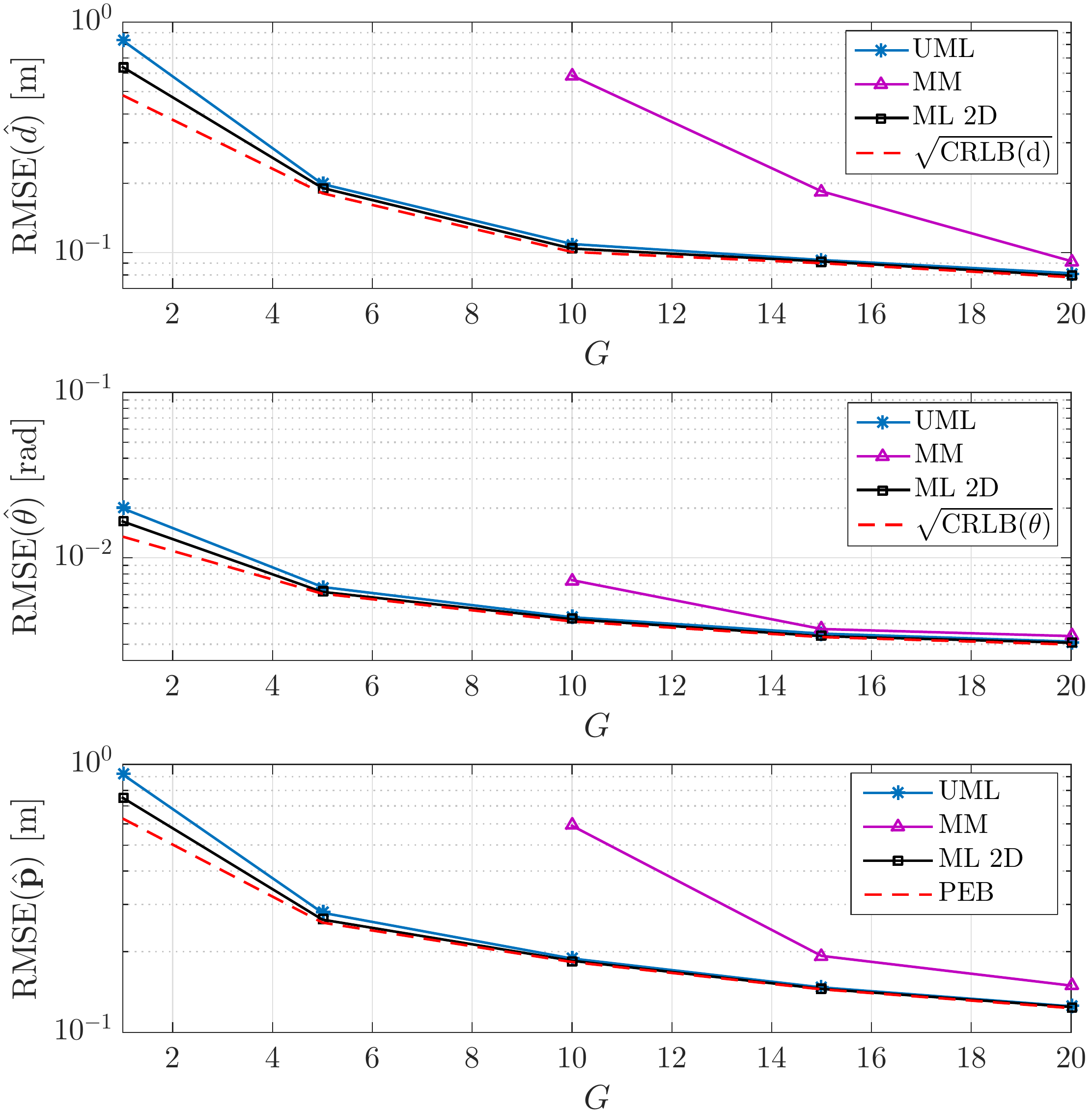}
 	\caption{RMSEs of the estimated $d$, $\theta$, and $\bm{p}$ versus CRLBs as function of the transmissions $G$ for SNR = 5 dB in LOS condition.}
\label{fig:fig1}
 \end{figure}
 
In Fig.~\ref{fig:fig1} we depict the RMSE of the estimated values of $d = c \tau$, $\theta$ and $\bm{p}$ as function of the number of transmissions $G$ for SNR = 5 dB, compared against the theoretical bounds derived based on the FIM analysis in Sec. \ref{sec::FIM}. More precisely, the values of $\mathrm{\sqrt{CRLB(\point)}}$ are obtained similarly to the PEB defined in \eqref{eq::PEB}, i.e., by inverting the FIM $\bm{J}_{\bm{\gamma}}$ from \eqref{eq::FIM}, choosing the corresponding diagonal entries and taking the square root. For comparison, we have also added the performance of the two-dimensional ML estimator in \eqref{eq::loglike_3}. As it can be seen, the UML algorithm converges to the corresponding bounds for almost all the considered values of transmissions $G$. Remarkably, it exhibits very good estimation performance even in case of $G = 1$, achieving the same accuracy of the two-dimensional ML estimator, but at a significantly reduced computational cost. 

Starting from a sufficient number of transmissions $G = 10$ --- which we recall is the limit condition for defining the right pseudoinverse $\bm{X}^{+}_i$ required in \eqref{eq::pseudoinv} --- we can observe that the performance of the MM estimator are worse than that of the UML, but still acceptable in terms of achieved accuracy. As the solid (magenta) curves show, the RMSE of $\hat{d}$ exhibits a small gap with respect to the theoretical lower bound, which however starts to decrease as more transmissions $G$ are available, thanks to an increasingly accurate estimation of the first-order moment in \eqref{eq:mean_zi}. On the other hand, the RMSE of $\hat{\theta}$ approaches the theoretical bounds for $G \geq 15$. As the bottom plot in Fig.~\ref{fig:fig1} shows, this results in an overall position error which is always below 60 cm and tends to decrease as $G$ increases, achieving performance very close to the one provided by the UML, but at the lower cost of a single one-dimensional search. 

For the implementation of the MM algorithm, we compared the one-lag based estimator in \eqref{eq::moment_tau} with the multi-lag extension in \eqref{eq::fritz}, assuming a number of lags $L = 2$. We observed that both estimators exhibit almost the same performance in the considered scenario.

\begin{figure}	
\centering
 \includegraphics[width=0.51\textwidth]{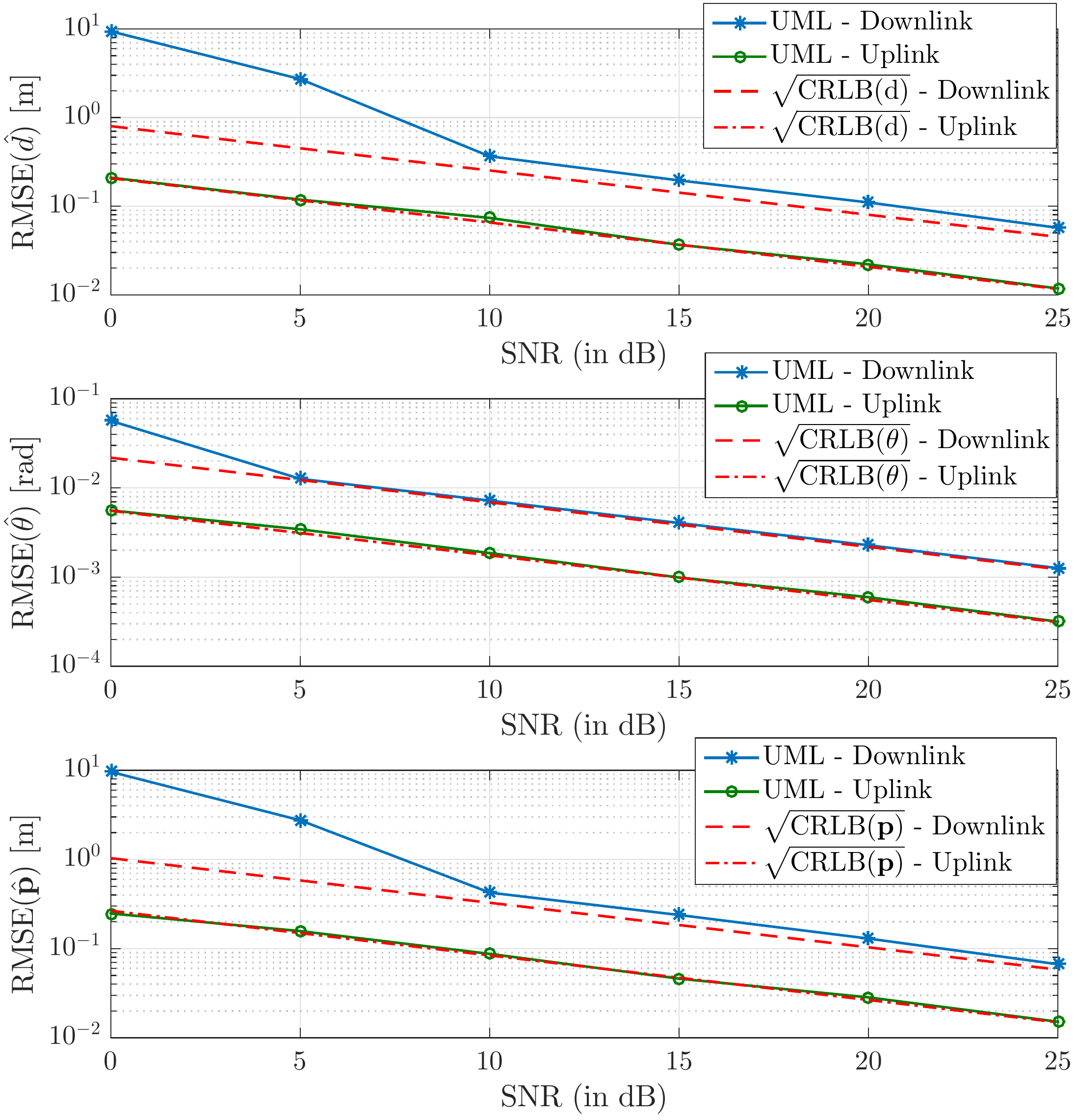}
 	\caption{RMSEs of the estimated $d$, $\theta$, and $\bm{p}$ versus CRLBs as function of the SNR for $G = 1$ in LOS condition for both uplink and downlink channels.}
\label{fig:fig2}
 \end{figure}
 
Fig.~\ref{fig:fig2} shows the evolution of the RMSEs of $d$, $\theta$ and $\bm{p}$ with respect to different values of the received SNR for the challenging case of $G = 1$ in both uplink and downlink channels. It is worth noting that, in this case, the pseudoinverse matrix $\bm{X}^{+}_i$ in \eqref{eq::pseudoinv} is not defined and hence the MM estimator cannot be implemented. 
By comparing the obtained results, it can be seen that a higher estimation accuracy is achieved in the uplink channel. This behavior can be linked to the fact that, in the uplink channel, a $N_{\text{BS}}$-dimensional vector of samples is available for each subcarrier, thus resulting in better estimation conditions with respect to the downlink channel, as confirmed by the smaller values of the bounds (dashed-dot red curves). 

Interestingly, the proposed UML estimator performs well even for very low values of the received SNR, which is a typical operating condition in mmWave systems before beamforming stage. On the other hand, it should be noticed that the UML  attains the theoretical bounds also in the downlink channel, starting from values of SNR of about 10 dB. 
%As a result, the performance gap in terms of position error between uplink and downlink channels is not dramatic and significantly decreases as the SNR increases, as shown in the bottom figure.
%both the suboptimal estimators of $\tau$ and $\theta$ can provide the same accuracy of the 2-dimensional ML estimator in \eqref{eq::loglike_3}, but at the reduced cost of few 1-dimensional searches.
%, that is, one for the estimation of $\theta$ and one for the refinement of $\tau$. 

\subsection{AOD, TOF and Position Estimation in LOS + NLOS}\label{subs::LOS_NLOS}
In this section, we evaluate the algorithms performance assuming that, in addition to the direct LOS path, two NLOS paths due to multipath propagation  are also present.
\begin{figure}	
 \centering
 \includegraphics[width=0.51\textwidth]{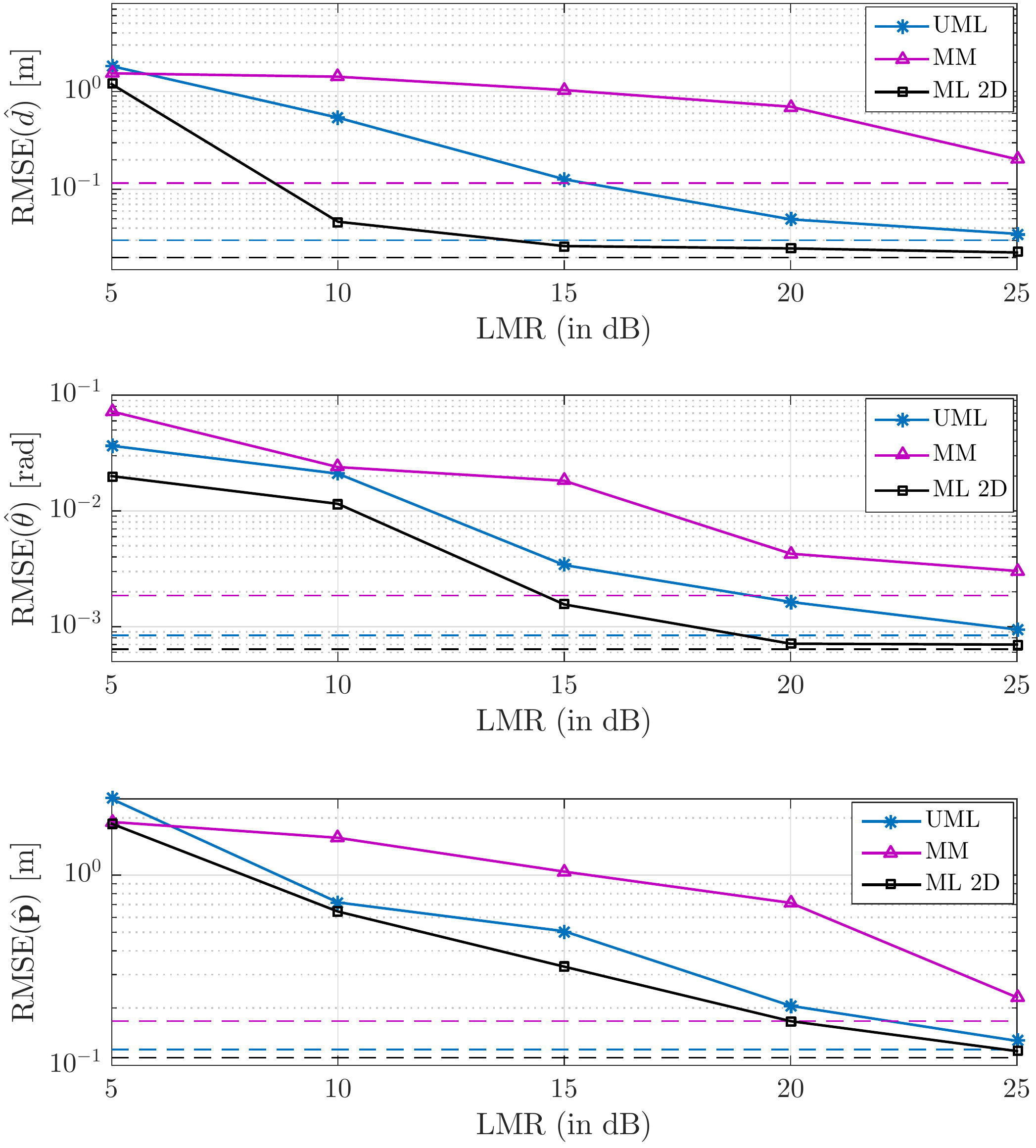}
 	\caption{RMSEs of the estimated $d$, $\theta$, and $\bm{p}$ as function of the LMR for $G = 10$ and SNR = 15 dB. The dashed curves represent LOS-only performance for reference.}
\label{fig:fig3}
 \end{figure}
 
In Fig.~\ref{fig:fig3} we report the RMSEs of $d$, $\theta$ and $\bm{p}$ as function of different levels of LMR for $G = 10$ and SNR = 15 dB.
As it could be expected, the obtained performance are slightly worse than that in LOS-only (depicted as dashed curves for reference), especially for non-negligible levels of multipath power. Nevertheless, a high level of position accuracy can still be achieved for mid to high values of LMR, with the UML algorithm exhibiting the same estimation performance of the more computationally demanding ML 2D. 

The localization capability of the MM algorithm is also interesting, at least for non-severe multipath conditions, as confirmed by the solid (magenta) curves. These results show that the proposed algorithms are effective also in presence of multipath propagation effects.

\section{Conclusion}\label{sec:conclusions}

We have addressed the problem of determining the unknown MS position under a mmWave MISO system setup. Our solution is based on the ML approach and exploits the AOD of received downlink signals, which can be conveniently estimated using a single receive antenna, thus providing an efficient way for locating a receiver while avoiding the high computational cost required by large arrays. We have performed a thorough theoretical analysis, providing the exact solution to the ML estimation problem and deriving the fundamental lower bounds on the estimation uncertainty for both channel and position parameters. To circumvent the need for multidimensional optimization to compute the joint ML estimator, we proposed two novel approaches amenable to practical implementation thanks to their reduced complexity. The performance assessment demonstrated that different accuracy/complexity trade-offs exist; however, remarkably, it is possible to achieve almost the same performance of the exact ML estimator, at a fraction of its computational complexity, even in presence of few transmissions, low SNRs, and multipath propagation effects.

\appendix
\subsection{Necessary conditions for non-singular FIM}
For clarity, we drop $\tau$ from the unknowns and consider the case
$G=1$ (allowing us to drop the index $g$), and introduce 
\begin{align*}
\zeta[n] & =\bm{a}^{\mathrm{H}}(\theta)\bm{s}[n]\\
\xi[n] & =\bm{a}^{\mathrm{H}}(\theta)\bm{B}\bm{s}[n],
\end{align*}
from which we create length $N$ vectors 
\begin{align}
\bm{\zeta}& =\bm{S}^{\mathrm{T}}\bm{a}^{*}(\theta)\\
\bm{\xi}&=\bm{S}^{\mathrm{T}}\bm{B}\bm{a}^{*}(\theta),
\end{align}
where $\bm{S}$ has as columns the $N$ pilots $\bm{s}[n]$.
The FIM is then given by (up to irrelevant constants)
\begin{align}
&\bm{J}_{\bm{\gamma}}=\\
& \left[\begin{array}{ccc}
\bm{\zeta}^{\mathrm{H}}\bm{\zeta} & 0 & r\pi \cos(\theta)\Im(\bm{\xi}^{\mathrm{H}}\bm{\zeta})\\
0 & r^2\pi \cos(\theta)\bm{\zeta}^{\mathrm{H}}\bm{\zeta} & r\pi \cos(\theta)\Re(\bm{\xi}^{\mathrm{H}}\bm{\zeta})\\
r\pi \cos(\theta)\Im(\bm{\xi}^{\mathrm{H}}\bm{\zeta}) & r\pi \cos(\theta)\Re(\bm{\xi}^{\mathrm{H}}\bm{\zeta}) & r^2\pi^2 \cos^2(\theta)\bm{\xi}^{\mathrm{H}}\bm{\xi}
\end{array}\right] \nonumber 
\end{align}
where $\Im\{\point\}$ denotes the imaginary-part operator. This matrix is full-rank if and only if its determinant is non-zero.
After some math, we find that 
\begin{align*}
\det\bm{J}_{\bm{\gamma}} & =0\\
\iff\\
\Vert\bm{\zeta}\Vert^{2}\Vert\bm{\xi}\Vert^{2} & =|\bm{\xi}^{\mathrm{H}}\bm{\zeta}|^{2}
\end{align*}
From the Cauchy-Schwarz inequality for complex numbers, it then follows
that the FIM is singular if and only if $\bm{\zeta}$ and $\bm{\xi}$
are parallel, i.e., there exists a $u\in\mathbb{C}$ so that $\bm{\xi}=u\bm{\zeta}$.
It then follows immediately that:
\begin{itemize}
\item when $N=1$, $\bm{\zeta}$ and $\bm{\xi}$ become scalars, which
are necessarily parallel and thus $\bm{J}_{\bm{\gamma}}$ is singular;
\item when $N>1$, a necessary and sufficient condition for a non-singular
$\bm{J}_{\bm{\gamma}}$ is that the columns of $\bm{S}$ are not
parallel. 
\end{itemize}
Thus, a necessary and sufficient condition for a non-singular FIM
with a single transmission ($G=1$) is that we use at least $N=2$
subcarriers and that the pilots across those subcarriers are different
(non parallel). It can similarly be shown that when $N=1$, we must
use at least two transmissions with different pilots. 

\ifCLASSOPTIONcaptionsoff
  \newpage
\fi

% % biography section
% % 
% % If you have an EPS/PDF photo (graphicx package needed) extra braces are
% % needed around the contents of the optional argument to biography to prevent
% % the LaTeX parser from getting confused when it sees the complicated
% % \includegraphics command within an optional argument. (You could create
% % your own custom macro containing the \includegraphics command to make things
% % simpler here.)
% %\begin{IEEEbiography}[{\includegraphics[width=1in,height=1.25in,clip,keepaspectratio]{mshell}}]{Michael Shell}
% % or if you just want to reserve a space for a photo:

% \begin{IEEEbiography}{Author1}
% Biography text here.
% \end{IEEEbiography}

% % if you will not have a photo at all:
% \begin{IEEEbiographynophoto}{Author2}
% Biography text here.
% \end{IEEEbiographynophoto}

% % insert where needed to balance the two columns on the last page with
% % biographies
% %\newpage

% \begin{IEEEbiographynophoto}{Author3}
% Biography text here.
% \end{IEEEbiographynophoto}

% % You can push biographies down or up by placing
% % a \vfill before or after them. The appropriate
% % use of \vfill depends on what kind of text is
% % on the last page and whether or not the columns
% % are being equalized.

% %\vfill

% % Can be used to pull up biographies so that the bottom of the last one
% % is flush with the other column.
% %\enlargethispage{-5in}

\end{document}